\documentclass[aps,prc,onecolumn,showkeys,showpacs,floatfix]{revtex4}
\usepackage{graphicx}
\usepackage{epsfig}
\usepackage{soul}
\usepackage{hyperref}
\usepackage{amsmath}

\begin{document}\title{Cross-sections for $\nu_{\mu}$ and $\bar{\nu_{\mu}}$ induced single pion production processes in the few GeV using NuWro}
\author{Qudsia Gani$^{1}$\footnote{Corresponding author:  ganiqudsia@kashmiruniversity.net} and W. Bari$^{2}$ }
\affiliation{$^1$ Department of Physics, Govt. College for Women, M.A. Road Srinagar, 190 001, India}            
\affiliation{$^2$ Department of Physics, University of Kashmir, Srinagar 190006, India}            

\begin{abstract}
\textbf{Abstract:} For a continued progress in the pursuit of neutrino oscillation parameters, we require to have a precise knowledge of the pion production processes which act as a source of backgrounds and systematic uncertainties in neutrino oscillation experiments because pions can mimic a final state electron (positron). Therefore a pion production principal interaction  may appear to be quasi-elastic if the pion is absorbed in the final state interactions. Charged-current single pion production processes induced by $\nu_{\mu}$ and $\bar{\nu_{\mu}}$ interactions with hydrocarbon target $(CH_{2})$ has been studied using NuWro event generator and the cross-sections so obtained have been compared with the results of Miner$\nu$A experiment.    
\end{abstract}
\maketitle
\newpage

\section*{Introduction} Advancing our understanding of neutrino oscillation parameters requires building a more complete picture of neutrino interactions. This poses a series of important theoretical and experimental challenges. In this context neutrino event generators are an interface between theory and experiment and these play a vital role in the study of neutrino interactions from conception of an experiment to the final physics publication. Some of the important neutrino event generators based on Monte Carlo simulations are NuWro \cite{1}, GENIE \cite{2}, NEUT \cite{3}, FLUKA \cite{4} and NUANCE \cite{5}. The purpose of the event generators is to evaluate the feasibility  of some proposed experiment by way of optimizing the detector design, analyzing the collected data samples and evaluating the systematic errors and therefore assessing its physics reach. The use of simulations allows us to examine more complex systems than we otherwise can do. As for example, it may look fairly simple to solve the equations which describe the interactions between two atoms  but  for hundreds or thousands of atoms, it is not so easy to  solve the same equations. With simulations, a large system can be sampled in a number of random configurations and that data can be used to describe the system as a whole. This makes the neutrino event generators impressively polymorphic tools. 
\paragraph*{\bf The NuWro neutrino event generator:}
 NuWro \cite{1}, the Monte Carlo generator used for present study  handles all important processes in neutrino-nucleus interactions as well as the hadronization due to deep inelastic scattering (DIS) and intra nuclear cascade. It is light weight but full featured and serves as a tool to assess the relevance of various theoretical models  being investigated currently \cite{6}. It is organized around the event structure which contains three vectors of particles viz; incoming, temporary and outgoing. It also contains a structure with all the parameters used and a set of boolean flags tagging the event as quasi-elastic (QE), resonance excited scattering (RES), deep inelastic scattering (DIS), coherent pion production (COH) and meson exchange current (MEC); each through both schemes of interaction viz; charged-current (CC) and neutral-current (NC). The input parameters are read at start-up from a text file and the events are stored in the ROOT tree file to simplify further analysis. The basic algorithms of NuWro follow the other known codes such as NEUT, NUANCE, NEUGEN/GENIE etc. In order to facilitate comparisons, NuWro allows running simulations choosing easily the values of parameters, sets of form factors, models of nucleus etc. The distinguished features of NuWro are: fine hadronization model \cite{7}, description of resonance region without Rein-Sehgal approach \cite{8}  and effective implementation of spectral function as an improvement with respect to Fermi gas model \cite{9}. NuWro is a generator of interactions only. The neutrino is selected according to information about the beam and the target is selected as (nucleus or free nucleon). This is followed by choosing a model of nucleus such as (Fermi gas, local density approximation, effective potential, spectral function) and the internuclear cascade is switched on. The neutrino interaction point is selected inside nucleus according to the nuclear matter density. All secondary hadrons propagate through nucleus and can interact with nucleons inside. At each point of their path it is decided if there was an interaction or not. This is done based on an effective cross-section model, using NuWro.
   \newline The main motivation of the authors of NuWro was to have tools to investigate the impact of nuclear effects on directly observable quantities with all the final state interactions included. Now, NuWro simulates all the essential interactions and it is possible to be used in the experiments. As for instance, it has been included in the ICARUS  experiment\cite{10} with the task of improving NUX+FLUKA code in the single pion production region.
 \section*{Different competing neutrino-interaction processes at low energies}
 As the neutrino-flavour oscillation probability is inversely proportional to neutrino energy, the quest for determination of neutrino oscillation parameters forces us to do neutrino physics in few GeV range of energy.
\begin{equation}
 P(\nu_{\alpha}\rightarrow\nu_{\beta})= sin^{2}(2\theta)sin^{2}(\frac{\Delta m^{2}L}{4E_{\nu}})
\end{equation}
In this energy region, the main reaction mechanisms are quasi-elastic scattering (QE)\cite{11}, resonance excited scattering (RES) \cite{12} and single pion production through  $\Delta$ excitation \cite{13}, each through both schemes of interaction viz: charged-current (CC) and neutral-current (NC).  The quasi-elastic process where there is only a muon and a nucleon in the final state, constitutes a large fraction of the signal population in the first few GeV and has been abundantly studied. 
In case of charged-current quasi-elastic interactions, with baryon targets, the baryon undergoes a change in its electric charge to accommodate the exchange of the charged \(W ^\pm\)  boson.
\begin{equation}
\nu_{\mu} + n \rightarrow  \mu^{-} + p
\end{equation}
 If, instead of simply altering the charge of the target baryon, the W$^{{\underline{+}}}$ boson transfers enough momentum to promote the target into a low-mass resonance state, then the decay of the resonance  typically produces a nucleon and a pion mainly through  resonant mechanisms. The neutrino and anti-neutrino scattering off the free nucleons comprises of seven resonant single pion reaction channels for each, which include three charged-current processes viz;
 \begin{equation}
\nu_{\mu}p \rightarrow\mu^{-}p\pi^{+} \hspace{2cm} \bar{\nu_{\mu}}p\rightarrow\mu^{+}p\pi^{-} 
\end{equation}
\begin{equation}
\nu_{\mu}n \rightarrow\mu^{-}p\pi^{0} \hspace{2cm} \bar{\nu_{\mu}}p\rightarrow\mu^{+}n\pi^{0}
\end{equation}
\begin{equation}
\nu_{\mu}n \rightarrow\mu^{-}n\pi^{+} \hspace{2cm} \bar{\nu_{\mu}}n\rightarrow \mu^{+}n\pi^{-}
\end{equation} 
and four neutral-current processes viz;
\begin{equation}
\nu_{\mu}p\rightarrow\nu_{\mu}p\pi^{0} \hspace{2cm}\bar{\nu_{\mu}}p\rightarrow\bar{\nu_{\mu}}p\pi^{0}
\end{equation}
\begin{equation}
\nu_{\mu}p\rightarrow\nu_{\mu}p\pi^{+} \hspace{2cm}\bar{\nu_{\mu}}p\rightarrow\bar{\nu_{\mu}}p\pi^{0}
\end{equation}
\begin{equation}
\nu_{\mu}n\rightarrow\nu_{\mu}n\pi^{0} \hspace{2cm}\bar{\nu_{\mu}}n\rightarrow\bar{\nu_{\mu}}n\pi^{0}
\end{equation}
 \begin{equation}
 \nu_{\mu}n\rightarrow\nu_{\mu}n\pi^{-} \hspace{2cm}\bar{\nu_{\mu}}n\rightarrow\bar{\nu_{\mu}}p\pi^{-}
\end{equation}
Depending on the neutrino energy, the single pion production processes through baryonic resonances created in neutrino-nucleon interactions can also potentially decay to multipion final states \cite{14}. 
 \newline Apart from the resonance excited scattering, single pion final states are also produced when the neutrino is coherently scattered off the entire nucleus, transferring negligible energy to the target say, (A). This can proceed through both neutral-current and charged-current  processes;
\begin{equation}
\nu_{\mu}  A \rightarrow  \nu_{\mu}  A \pi^{0}
\hspace{2cm} \nu_{\mu}  A \rightarrow  \mu^{-}  A \pi^{+}
\end{equation}
Such processes are referred to as coherent pion production (COH).
\newline Finally, deep inelastic scattering (DIS) events can also contribute a copious source of neutral pions in the final states which may contaminate an electron-neutrino appearance measurement \cite{15}. 
\newline  In general QE processes are characterised by topologies with no pions in the initial and final states whereas DIS and RES processes result in events with pions in the primary and final state.
\begin{figure}[h]
\centerline{\includegraphics[trim=0cm 0cm 0cm
0cm,width=0.35\textwidth,clip]{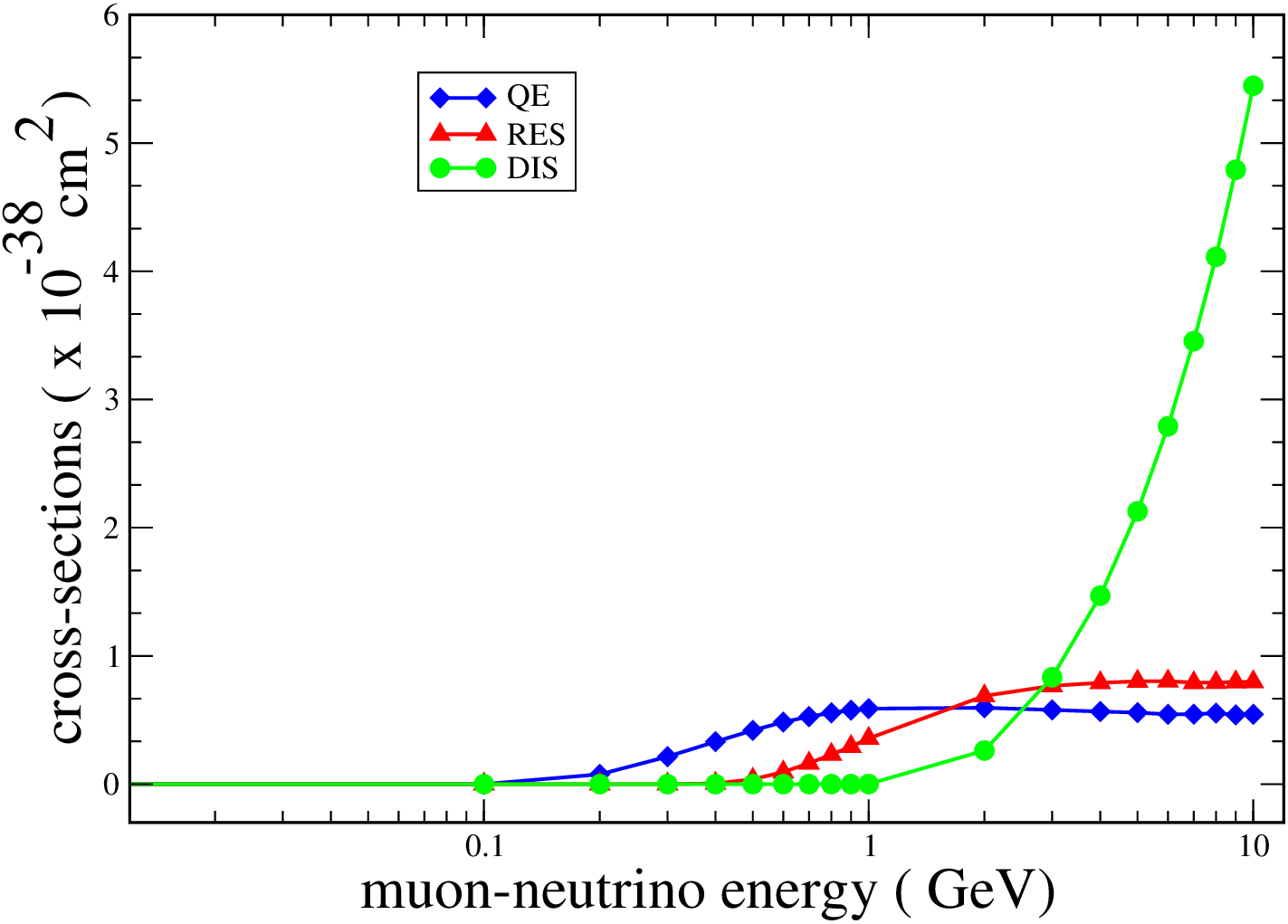}} \caption{cross-sections with (water) using NuWro} \label{figure 1}
\end{figure}\\ 
At $\sim$ 1 GeV  of neutrino energy, the charged-current resonant single pion production accounts for the second largest contribution
to the  cross-section after charged-current quasi-elastic (CCQE) scattering. However the effect of the two processes gets mingled when a pion production process is mistaken for quasi-elastic signal if the pion is later absorbed in the target nucleus. Therefore, the experiments which reconstruct neutrino energies through quasi-elastic events can yield an incorrect estimate of the incident neutrino energies.  Moreover, as almost all the generators have a larger number of zero pion topologies in the final state than were in the primary state, this also indicates that pions are more likely to be absorbed than created. Depending on cuts, this effect has been shown to account for about 10-20 $\%$ of quasi-elastic events \cite{16}. Therefore an enhanced cross-sections of QE processes as shown by Fig.1 above at low energies may actually carry some illusions. However, it is pertinent to mention that the linearly rising cross-sections due to QE and RES processes also get damped by the nuclear form factors as the neutrino energy increases. 
 \newline Thus, when neutrino interactions take place in the nucleus, the secondary particles or the hadrons that are produced, can interact with the nuclear medium and modify the observed characteristics of the interactions.  In this context, the production of pions induced by neutrinos is of much interest because pions are particularly susceptible to the effects of the nuclear medium, since these interact via the strong nuclear force. Charged pions are either absorbed or get converted into neutral pions via:
\begin{equation}
n +  \pi^{+} \longrightarrow p +  \pi^{0}
\end{equation}  
 The nuclear medium can also influence whether a pion is even created. All these factors have a direct bearing on the neutrino interaction cross-sections. 
\section*{Cross-section for pion production processes}
   The charged-current single pion production processes are especially important in the few GeV range of neutrino energy in long baseline neutrino oscillation experiments \cite{17}.  In this study, we have generated the cross-sections for $\nu_{\mu}$ and $\bar{\nu_{\mu}}$ interactions with hydrocarbon target using NuWro event generator. The topology selections in this study are restricted to resonance excited scattering (RES) and coherent pion production (COH) only whileas the contributions from quasi-elastic (QE) and deep inelastic scattering (DIS) processes are negligible. \newline Fig. 2 and Fig. 3 show the NuWro predictions for RES processes producing pions through charged-current $(\nu/\bar{\nu})$ interactions.
  \begin{figure}[h]
\centering
\begin{minipage}{0.49\textwidth}
\centering
\includegraphics[scale=.28]{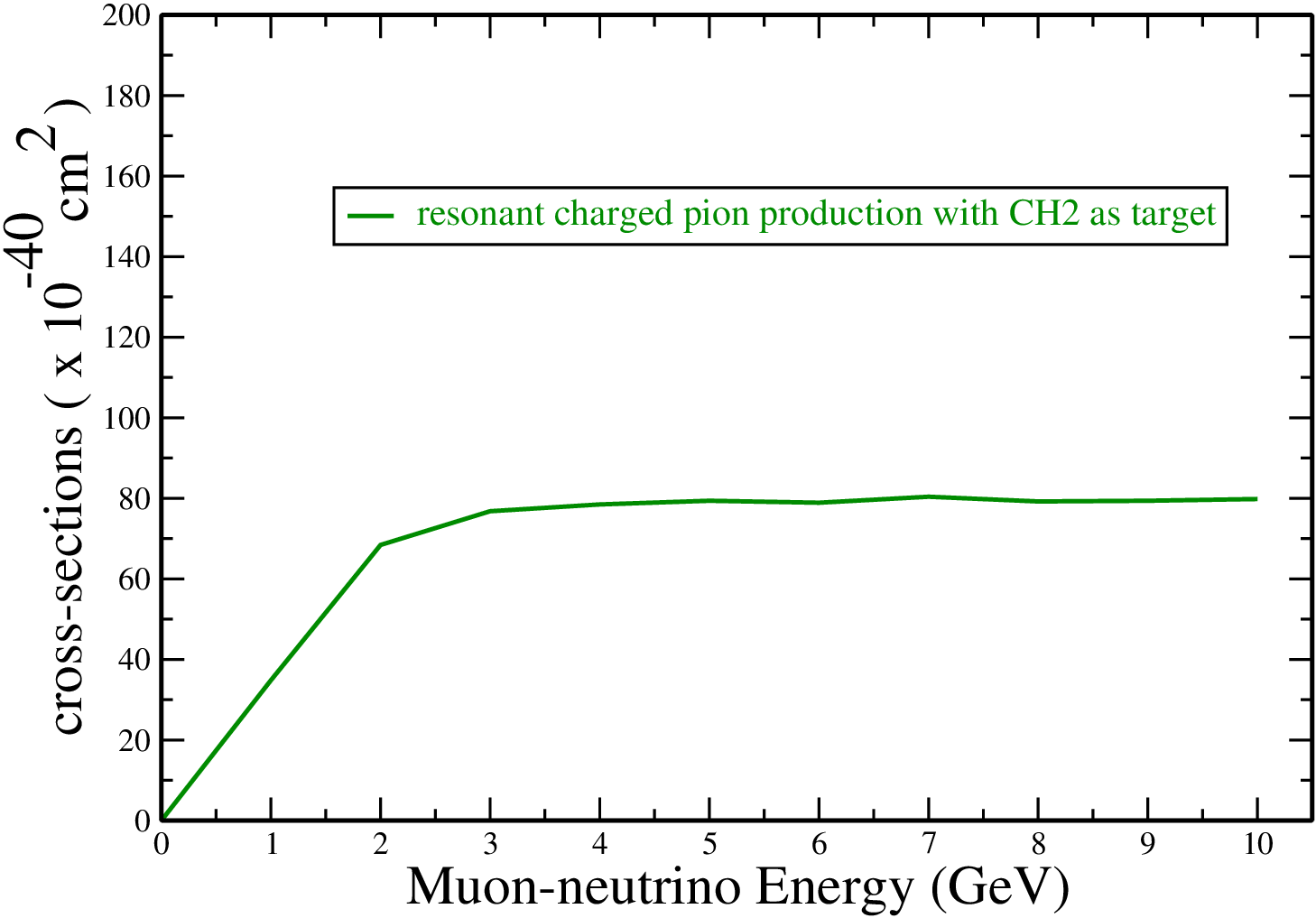} \caption{$\nu_{\mu} + CH_{2} \rightarrow \mu^{-} + \pi^{+} + X$} \label{figure 2}
\end{minipage}
\begin{minipage}{0.49\textwidth}
\centering
\includegraphics[scale=.28]{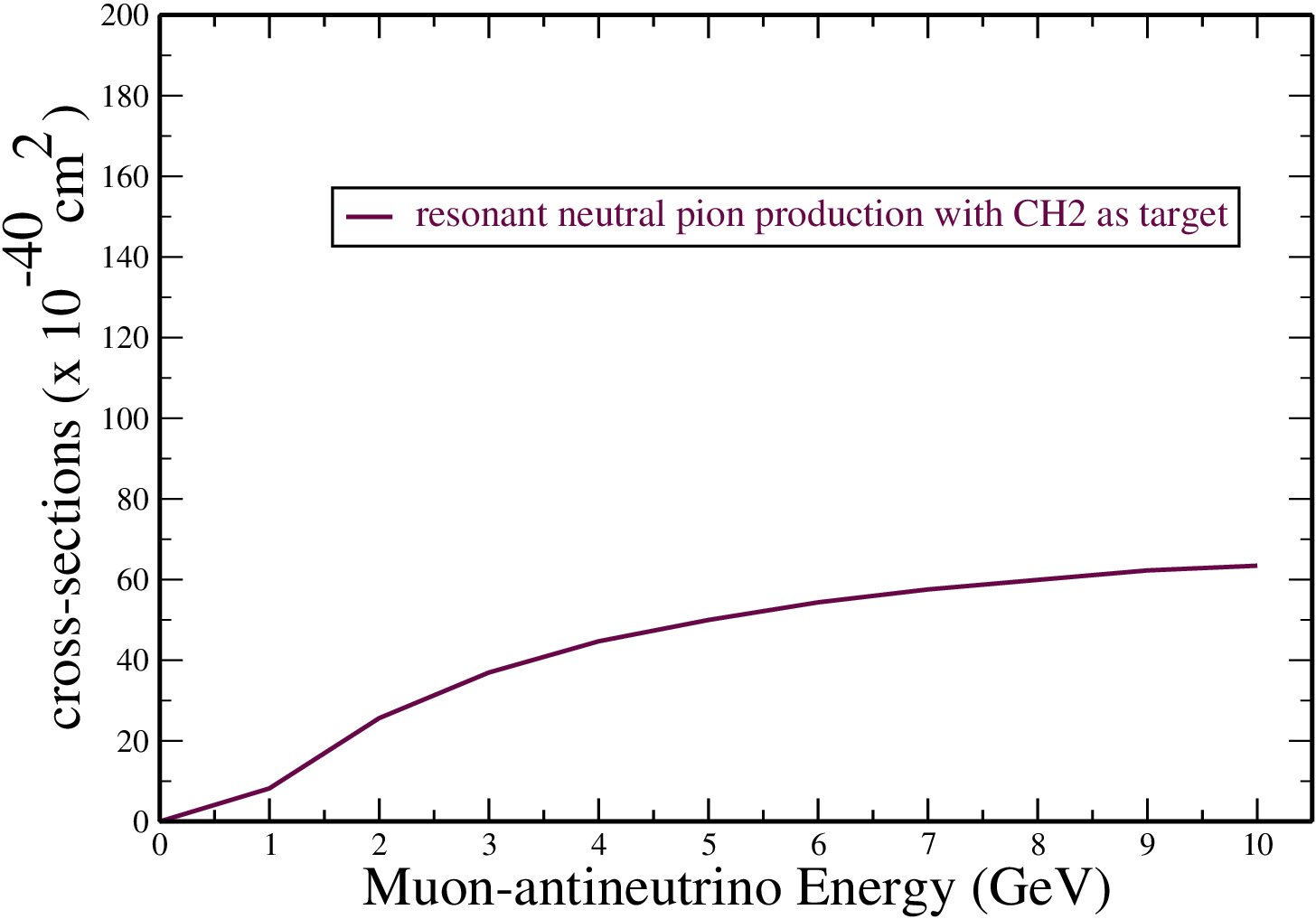} \caption{$\bar{\nu_{\mu}} + CH_{2} \rightarrow \mu^{+} + \pi^{o} + X$} \label{figure 3}
\end{minipage}
\end{figure}\\
The cross-section for $\nu_{\mu} CC (\pi^{+})$ is nearly twice as the $\bar{\nu_{\mu}} CC (\pi^{o})$ cross-section. Whileas the $\nu_{\mu} CC (\pi^{+})$ process attains its plateau at $E_{\nu}$ = 3 GeV, the $\bar{\nu_{\mu}} CC (\pi^{o})$ process shows a gradual rise throughout the measured region. This behaviour is largely the manifestation of underlying vector minus axial-vector (V-A) structure of the hadronic currents which interfere constructively in $\nu_{\mu} CC (\pi^{+})$ process and destructively in $\bar{\nu_{\mu}} CC (\pi^{o})$ process. These interferences have a significant contribution to the cross-sections in hundreds of MeV to few GeV range of neutrino energy. The predictions of NuWro exceed slightly the measured $\nu_{\mu} CC (\pi^{+})$ cross-section by MINER$\nu$A \cite{18} but is still much better than GENIE and NEUT which exhibit a much larger disagreement. However, in case of $\bar{\nu_{\mu}} CC (\pi^{o})$ cross-section, there is a much better agreement with the data and also a less variation among the generators \cite{18}.
\newline For coherent pion production, as it is required that the muon-pion system should be chargeless as the incident neutrino/antineutrino, the production of neutral pion cannot occur through this process. It proceeds through $\nu_{\mu} CC (\pi^{+})$ mode only, the cross-section of which is shown in Fig. 4.  
\begin{figure}[h]
\centerline{\includegraphics[trim=0cm 0cm 0cm
0cm,width=0.35\textwidth,clip]{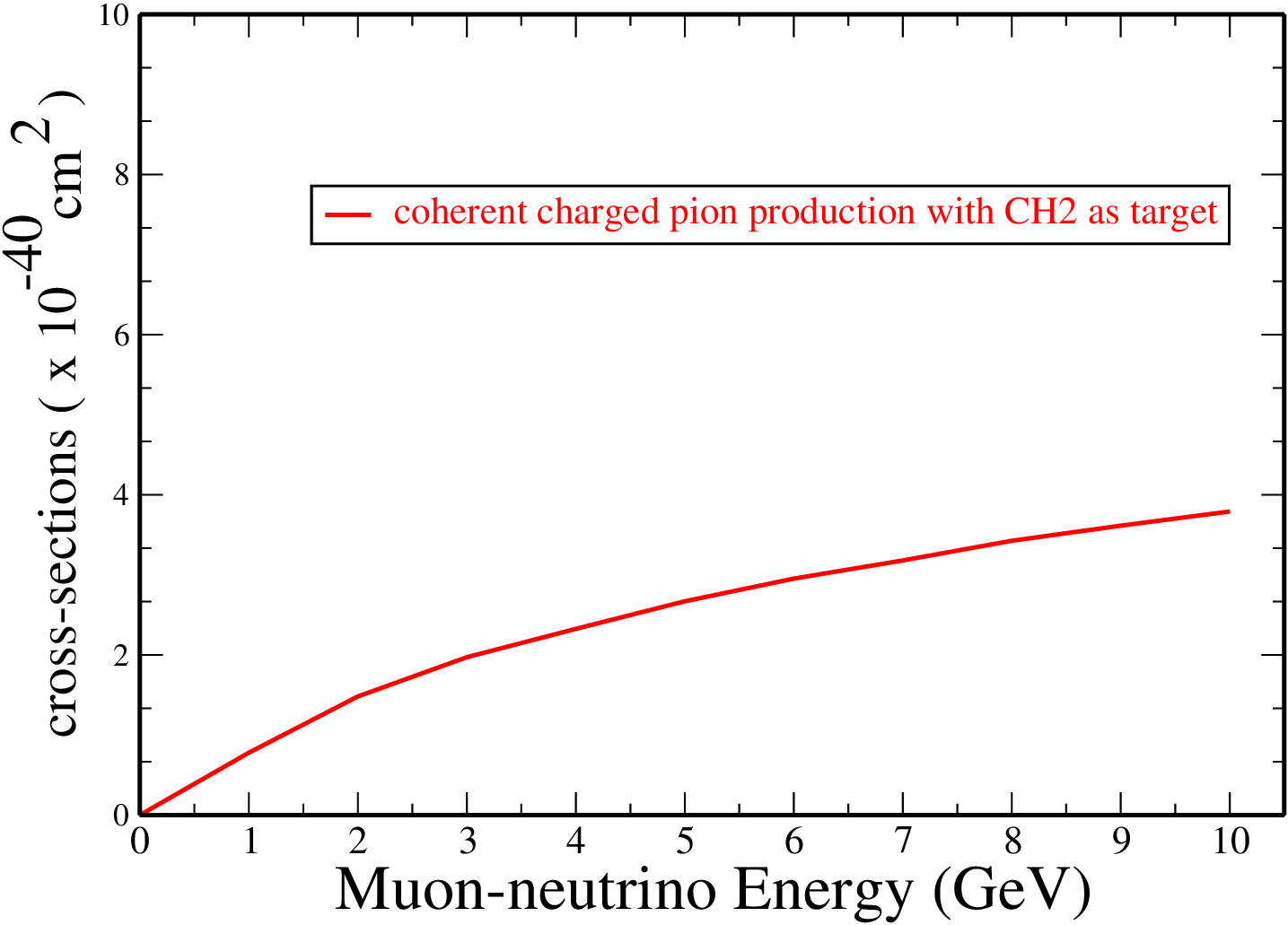}} \caption{$\nu_{\mu} CC (\pi^{+})$} \label{figure 4}
\end{figure}\\
When the  plot  in Fig. 4 is compared with those in Fig. 2 and Fig. 3, the clear inference is that $\Delta(1232)$ resonance due to RES dominates at low $E_{\nu}$. However, its contribution is expected to fall  at higher energies where the struck nucleon may further get excited. The separation of  resonant and other processes is model dependent and could be different in different generators.
\newline The next question is if we should  focus on pion-nucleus cross-sections and how useful is the pion transparency data?  
 For the effective measurement of pion-nucleon cross-sections, the NuWro FSI code has recently been incorporated with the Oset model \cite{19}. The NuWro predictions are obtained in the standard way by arranging a homogeneous flux of pions and counting the particles in the final state assuming that at least one microscopic interaction took place. The sum over all possible interaction channels gives the pion-nucleus reaction cross-section which we intend to take up in next studies.     
\section*{Summary}
One of the most straightforward approaches to improve the understanding of neutrino oscillations is to investigate the various processes contributing to the total cross-sections in neutrino-nucleus interactions. Keeping in view the importance of such studies, an attempt has been made in the present contribution to measure the cross-sections for  pion production processes. Because of the inherent complexity of reconstructing multiple pion final states, there are not many existing experimental measurements of such process. Therefore we have tried to restrict the present study to single pion production processes only.  Our results have been obtained using the NuWro event generator in some particular user-defined situations where muon-neutrinos/antineutrinos are scattered off the fixed nuclear targets like water and hydrocarbon. The same are substantiated by the results of MINER$\nu$A experiment. 
 While the results do not improve on precision, these serve as a useful cross check in a region with few measurements and also verify the  experimental measurements. 
\section*{Acknowledgements}
  The authors are highly grateful to J. Sobczyk of the Institute of Theoretical Physics, Wroclaw, Poland for fruitful communications with him from time to time which lead to initiate this type of study at University of Kashmir.
  \paragraph*{}
  \paragraph*{}
  \paragraph*{}The authors declare that there is no conflict of interest regarding the publication of this paper.
  \section*{References}
\begin{itemize}
\bibitem{1} T. Golan, C. Juszczak,  J. T. Sobczyk;  \textit{Phys. Rev. C} {\bf 86},   015505 (2012).
\bibitem{2}J.Dobson, C.Andreopoulos;  \textit{Acta Phys.Pol. B} {\bf 40}, 2613 (2009).
\bibitem{3}M. Nakahata et al;  \textit{J.Phys. Soc. Jps.} {\bf 55}, 3786 (1986).
\bibitem{4}A.Ferrari, P.R. Sala, A. Fasso, J. Ranft,  CERN-2005-010, INFN/TC-05/11 (2005)
\bibitem{5}D.Casper, The Nuance neutrino physics simulation and the future
http://nuint.ps.uci.edu/nuance/files/nuance-nuint01.pdf
\bibitem{6} J.A. Nowak,  \textit{Phys.Scr. T} {\bf 127}, 70 (2006).
\bibitem{7} K. S. Kuzmin, V. V. Lyubushkin,  V. A. Naumov,  Phys. Atom. Nucl. {\bf 69}, 1857 (2006).
\bibitem {8} D. Rein and L.M. Sehgal; \textit{Nucl. Phys. B} {\bf 223}, 29 (1983).
\bibitem{9}Gerald A. Miller; Fermi Gas Model; \textit{Nuclear Physics B}  \textbf{112},  223–225 (2002).
 \bibitem{10}  ICARUS Collaboration, \textit{Acta Phys. Polon B } \textbf{41}, 103-125 (2010). 
\bibitem{11}  Jan T. Sobczyk $arXiv:1108.0506v1 [hep-ph]$ (2011).
\bibitem{12} K. M. Graczyk, D. Kielczewska, P. Przewlocki, J. T. Sobczyk; \textit{Phys. Rev. D} { \bf 80}, 093001 (2009).
\bibitem{13} Krzysztof M. Graczyk, Jakub Żmuda, and Jan T. Sobczyk;  \textit{Phys. Rev. D} \textbf{90}, 093001 (2014).
\bibitem{14} J Formaggio,  and G Zeller  \textit{Reviews of Modern Physics} {\bf 84.3 }1307–1341 (2012).
\bibitem{15} Deborah A. Harris, $arXiv:1506.02748v1 [hep-ex]$ (2015).
\bibitem{16}  T. Leitner, \textit{Phys. Rev. C} \textbf{81}, 064614 (2010).
\bibitem{17} T. Le et al; \textit{Phys. Lett. B} \textbf{749}, 130-136 (2015).
\bibitem{18}C. L. McGivern et al; \textit{Phys. Rev. D} \textbf{94}, 052005 (2016).
\bibitem{19}L. L. Salcedo, E. Oset, M. J. Vicente-Vacas, and C. Garcia-Recio;  Nucl. Phys. A {\bf 484}, 557 (1988).

\end{itemize}

\end{document}